\documentclass[12pt]{article}
\usepackage{amsmath}
\usepackage{amsfonts}
\usepackage{amssymb}
\usepackage{graphicx}

\input{tcilatex}
\begin{document}

\begin{titlepage}
\ \\
\begin{center}
\LARGE
{\bf
Secure Quantum Passwords
}
\end{center}
\ \\
\begin{center}
\large{
Masahiro Hotta${}^{\ast}$
\footnote[1]{hotta@tuhep.phys.tohoku.ac.jp}
and Masanao Ozawa${}^{\S}$
\footnote[2]{ozawa@math.cm.is.nogoya-u.ac.jp}
}\\
\ \\
\ \\
{\it
${}^\ast$
Graduate School of Science, Tohoku University,\\
Sendai, 980-8578,   Japan}\\
\ \\
{\it 
${}^{\S}$
Graduate School of Information Science, Nagoya University,\\
Chikusa-ku, Nagoya, 464-8601,  Japan }
\end{center}
\begin{abstract}
We propose a quantum authentication protocol that is robust against the theft of secret keys. 
In the protocol, disposable quantum passwords prevent impersonation attacks with stolen secret keys. 
The protocol also prevents the leakage of  secret information of a certification agent.
\end{abstract}
\end{titlepage}

\section{Introduction}

\ \newline

Secure authentication plays an important role in modern society, \
supporting various types of transactions. Recently, there has been a surge
in crimes involving skimming passwords in smart cards. In addition, computer
viruses and criminals have led to authentication systems leaking customer
passwords from certification agents which should be kept secret under normal
circumstances. These issues expose the fragility of classical authentication
methods. \ Classical authentication is performed essentially by
cross-checking classical secret information composed of alphanumeric
characters and shared by two parties to identify each other. Although
improved methods have been developed which apply biometrics including
fingerprints and iris codes, these cannot stop skimming at a fundamental
level. The root of the problem is that there exists no principle to prohibit
the cloning of classical information.

Meanwhile, quantum authentication is a possible way to assure such safety by
employing fundamental physics principles. Under the no-cloning theorem \cite%
{wz} of quantum information, there exists no physical process to perfectly
copy quantum states non-orthogonal to each other. Hence, by encoding secret
information into non-orthogonal quantum states, it is possible to prevent
perfect skimming. It is also known from the uncertainty principle \cite{H}
that measurements for eavesdropping by an adversary can be detected by
checking the change of quantum states.

Several quantum protocols for identification have been proposed. Those
methods can be classified into two basic classes of method. In the first
method, common secret keys are generated as classical information composed
of alphanumeric characters \cite{c-key}-\cite{Guo}. In identification
processes, shared classical information is converted into quantum
information and sent from $A$, who is a user, to $B$, who is a certification
agent. In the second method \cite{B}--\cite{CS}, secret keys are shared as
quantum information from the start. For example, $A$ stores the information
in a specific portable device, a quantum smart card, which is slotted into
the quantum authentication machines of $B$. Though these quantum protocols
have advantages over classical methods, several unsatisfactory aspects
remain. For example, in the first method, clone leakage of classical secret
keys from $B$ cannot be prohibited in principle. For the second method,
impersonation cannot be prohibited when $A$'s quantum smart card is stolen.

In this paper, we propose a quantum protocol in which $A$ can require $B$ to
identify herself under high security, overcoming the above disadvantages by
using Bell states and one-time-pad passwords together. $A$'s $\,$qubits of
shared Bell pairs are stored in a quantum smart card. The password that $A$
determines is memorized only by $A$ and is not known by $B$ or others. The
advantages of our protocol are the following. (I) Even if $A$'s quantum
smart card is stolen by an adversary $E$, $E$ cannot impersonate $A$ without
knowing her password. (II) No available information of the password is
contained in the stolen card. (III) $B$ does not keep any information about $%
A$'s passwords, thereby avoiding the risk of clone-leakage from $B$'s
storage by $E$. (IV) $E$ cannot make perfect copies of the passwords by
eavesdropping on quantum channels. Moreover, by hiding quantum-mechanically
encoded passwords behind many decoy qubits when sending to $B$, a high rate
of eavesdropping detection can be achieved. (V) In identification tests of $%
A~$by $B$, entanglement between qubits of shared Bell pairs and qubits of \
passwords is not generated. Therefore, $B$ can use \ $A$'s quantum-password
qubits once and then throw them away. Hence, $B$ must resend to $A$ only
half of the qubits of the shared Bell pairs.

This paper is organized as follows. In section 2, typical protocols of
quantum authentication proposed to date are briefly reviewed. In section 3,
we propose a simplified protocol with quantum passwords by which we will
explain the basic ideas of our full protocol. In section 4, security
analysis is given for the simplified protocol. In section 5, an improved
protocol with high security is proposed by extending the simplified protocol
in section 3.

\bigskip

\section{Protocols proposed to date}

\bigskip

In this section, we characterize authentication by the following. (a) There
exist two parties $A$ and $B$. (b) The purpose of authentication is that $B$
identifies $A$ with high success probability. (c) $A$ and $B$ have common
secret keys that may be classical or quantum. (d) Authentication protocol
are composed of the following phases. (d,i) Generation of secret keys.
(d,ii) $A$ and $B$ work upon those secret keys by local operations and
individually store them. (d,iii) In authentication, $A$ and $B$ are able to
send their secret keys and other information using classical and quantum
communication. (d,iv) $B$ can check whether a user communicating with $B$ is
a legitimate person, that is $A$, using locally accessible information. If
something is wrong with this check, the process stops. (d,v) After $B$
recognizes $A$, they are able to perform local operations and exchange
information by classical and quantum communication. After that, \ the setup
of (d,ii) is reproduced.

If $A$ also wants to identify~$B$, the above protocol applies with exchanged
roles. The protocol is called quantum when it requires quantum information
and quantum media. In the following, we briefly review some proposed
protocols.

\textit{Classical authentication}: In (c) above, secret keys can be composed
of bit numbers. In (d,i), $A$ and $B$ meet and share a sequence of numbers.
In (d,ii), they individually store the numbers without revealing them to any
third party. For example, $A$ memorizes the secret key and $B$ stores the
key in an electronic database. In (d,iii), $A$ encodes the memorized
information by classical ciphers and sends it to $B$. $B$ decodes the
information sent from $A$. In (d,iv), $B$ compares the decoded result with
her sequence of secret numbers. If the decoded information is consistent
with the numbers, $B$ recognizes $A$. If not, $B$ stops the process. In this
protocol, there is no need for phase (d,v) because $B$ can discard the
information sent from $A$. It is well known that this protocol cannot
eliminate the danger of undetected leakage by cloning the classical
information.

\textit{Barnum 1}: Barnum \cite{B} proposed two quantum protocols. The first
method uses a sequence of qubit pairs in a fixed Bell state as secret keys
of phase (c). In (d,i), $A$ and $B$ each share half of the Bell pairs. In
(d,ii), $A$ stores her qubit pairs in a quantum smart card, while $B$ keeps
hers in a quantum storage device. In (d,iii), $A$ sends the quantum states
stored in the smart card to $B$ through a quantum channel. In (d,iv), $B$
performs a Bell measurement of the qubits sent from $A$ and the qubits
stored in $B$'s storage device. From the measurement results, $B$ verifies
whether the two qubit states in the sequence are an original Bell state. In
general, entanglement states such as Bell states exhibit purity properties
only when all the entangled subsystems are gathered. A lack of some
entangled subsystems leads to mixed states. If $A$ is a legitimate person,
the two qubits measured by $B$ should be in a pure Bell state to give an
acceptable result. If an adversary $E$ attempts to impersonate $A$, \ qubits
sent from $E$ cannot reproduce pure states with the qubits stored in $B$'s
storage device. Thus the Bell measurement by $B$ detects inevitable errors
for $E$'s qubits. In (d,v), after $B$ recognizes $A$ correctly, $B$ resends
half of the qubits of the Bell pairs in the right state to $A$. $A$ then
stores them again in her smart card. In this protocol, there is no danger of
clone leakage of $A$'s secret information from $B$, because $B$'s contracted
quantum state of qubits is the maximal entropy state. This is a remarkable
advantage over classical protocols. However, if $A$'s smart card is stolen
by $E$, $E$ is able to impersonate $A$ easily and the protocol becomes
insecure.

\textit{Barnum 2}: The second protocol proposed by Barnum \cite{B} uses
catalyst states. First, $A$ and $B$ share an entangled state $|\phi
_{1}\rangle $ of two qubits. \ We consider another state $|\phi _{2}\rangle $
of two qubits and assume that $|\phi _{2}\rangle $ is a state which is never
converted from $|\phi _{1}\rangle $ by local operations and classical
communication (LOCC). However, it is assumed to be possible that if $A$ and $%
B$ also share a certain entangled state $|\chi \rangle $, called a catalyst
state, $A$ and $B$ are able to transform $|\phi _{1}\rangle $ into $|\phi
_{2}\rangle $ by LOCC. Under the transformation, $|\chi \rangle $ remains
unchanged. An explicit setting of $|\phi _{1}\rangle ,|\phi _{2}\rangle $
and $|\chi \rangle $ can be seen in a standard textbook \cite{nc}. Using a
notion of catalyst states, Barnum proposed a quantum authentication protocol
as follows. In (d,i), $A$ and $B$ share a sequence of qubit pairs in a
catalyst state $|\chi \rangle $. In (d,ii), $A$ stores half of the qubits in
a quantum smart card and $B$ keeps the other half in a quantum storage
device. In (d,iii), $B$ generates a sequence of qubit pairs in $|\phi
_{1}\rangle $ and sends half of the pairs to $A$. In this step, \thinspace $%
A $ and $B$ share a sequence of four-qubit composite systems in a state $%
|\phi _{1}\rangle \otimes |\chi \rangle $. They transform $|\phi _{1}\rangle
\otimes |\chi \rangle $ into $|\phi _{2}\rangle \otimes |\chi \rangle $ by
LOCC. Then, $A$ resends half of the qubit pairs in $|\phi _{2}\rangle $ to $%
B $. In (d,iv), $B$ performs a measurement which verifies whether the \
composite systems of qubits sent from $A$ and the qubits stored by $B$
generated first in $|\phi _{1}\rangle $ are really in $|\phi _{2}\rangle $.
If $B$ gets a positive result, $B$ recognizes $A$. If not, $B$ stops the
process. When $E$ personates $A$, the transformation from $|\phi _{1}\rangle 
$ to $|\phi _{2}\rangle $ cannot be achieved by $E$ due to a lack of $|\chi
\rangle $. The advantage of this protocol is that there is no need for $A$
to send secret qubits in the catalyst state $|\chi \rangle $ to $B$. This
reduces the risk that the secret key could be stolen in the transfer from
\thinspace $A$ to $B$. However, as for \textit{Barnum 1,} if $A$'s smart
card, which contains the catalyst-state qubits, is stolen by $E$, $E$ can
impersonate $A$ easily.

\textit{Guo et al.}: In addition to the protocol of Barnum \cite{B}, a
method of quantum authentication by a different use of Bell states was
proposed by Guo et al. \cite{Guo}. In (d,i), $A$ and $B$ determine a
classical password as a sequence of numbers composed of \thinspace $1,2,3,4$%
. Each number is assigned to one of four orthogonal Bell states of two
qubits. This generates a sequence of Bell states along the order of \ the
classical password. In (d,ii), $A$ stores half of the Bell pairs in a
quantum smart card. $B$ keeps the other half in a quantum storage device and
also stores the classical password in an electronic database. In (d,iii), $A$
sends the qubits stored in the smart card to $B$. In (d,iv), $B$ performs a
Bell measurement of the qubits sent from $A$ and the qubits stored by $B$ so
as to read out the classical password. $B$ checks whether the obtained
results agree with the password stored in the database. If the results are
correct, \thinspace $B$ recognizes $A$. If not, $B$ stops the process. In
(d,v), $B$ resends half the Bell pairs to $A$. In this protocol, similar to 
\textit{Barnum 1}, the smart card does not contain any information about the
classical password because the contracted quantum states of the qubits in
the smart card are the maximal entropy state. However, this protocol is
insecure against card theft. Moreover, there is a risk that copies of the
classical password from $B$'s database could be leaked.

In the next section, we propose a secure protocol, which is robust against
card theft. The protocol also prevents information leak from the verifier $B$%
.

\bigskip

\section{Basic Protocol}

\bigskip

In this section, we explain a basic protocol in order to outline the essence
of the idea behind our full, more secure protocol, which we will be detail
in section 5. The structure of the basic protocol is composed as follows. In
(c) of the previous section, the secret keys are a sequence of qubit pairs
in a Bell state. In (d,i), $A$ and $B$ meet and generate a sequence of qubit
pairs in a Bell state $|+\rangle $. In (d,ii), $B$ puts half of the Bell
pairs into a quantum storage device, while $A$ stores the other half in a
quantum smart card. $A$ also generates a classical password composed of bit
values $0$ and $1$ with the same length of the above sequence of Bell pairs.
The password is not revealed to anybody, including $B$. $A$ performs a
unitary transformation on each qubit in the smart card, dependent on the bit
value of $A$'s password. When the bit value is $0$, the unitary
transformation is the identity transformation $I$. When the bit value is $1$%
, the unitary transformation is $R$, which is not the identity
transformation. The action of $R$ changes $|+\rangle $ into another Bell
state $|\xi \rangle $. The state $|\xi \rangle $ is not orthogonal to $%
|+\rangle $. In (d,iii), $A$ encodes the classical password by using two
non-orthogonal quantum states $|0\rangle $ and $|\alpha \rangle $ of a
qubit. The part of the password with bit values $0$ is replaced by the state 
$|0\rangle $. The other part with bit values $1$ is replaced by $|\alpha
\rangle $. We call the sequence of these qubits the quantum password. $A$
sends to $B$ both her quantum password and the qubits stored in her smart
card. In (d,iv), \thinspace $B$ combines the qubits sent by $A$ and the
qubits that $B$ keeps. The system becomes a sequence of composite three
qubits which contains a qubit of $A$'s quantum password and two qubits of
Bell pairs as secret keys. $B$ performs a unitary transformation $U$ on each
three-qubit system. As seen below, if the input of $U$ is legitimate, all
the states of the Bell-pair part become $|+\rangle $ independent of the bit
values of $A$'s password. \thinspace If not, other Bell components
orthogonal to $|+\rangle $ appear in the output of $U$ with nonzero
probability and give an error. $B$ performs a measurement to check whether
the state is really $|+\rangle $. If the result is positive, $B$ recognizes $%
A$. If not, $B$ stops the process. In (d,v), $B$ performs the inverse
transformation $U^{-1}$ to each set of three qubits in the sequence and
resends only half of the Bell-pair part to $A$. The other half is entered
again into $B$'s storage device. The qubits of $A$'s quantum password are
discarded by $B$ so that the information cannot be leaked.

In the following, we explain the protocol in more detail. First of all, we
specify the security levels of the environment. The region of $A$ is assumed
to be secure against any attack by an adversary $E$. Meanwhile, $E$ is
allowed to take classical information from the region of $B$, though $E$
cannot get any quantum media in or out the $B$'s region. This assumption
about $B$'s region implies for example that $E$ can steal the data of the
measurement results of $B$, but is not able to bring entangled qubits into $%
B $'s region for teleportation or to remove any quantum state of $B$. We
also assume that a public channel of classical communication is available
between $A$ and $B$ in which radiowaves with signals of $A$ spread widely in
open space towards $B$ and no adversary stops the communication. The channel
is used for announcements to $B$ of the start of $A$'s protocol.

We now give a detailed explanation using a password example. The protocol is
composed of nine steps, as follows.

\bigskip

(1) $A$ and $B$ meet and generate $N$ qubit pairs in a Bell state $|+\rangle 
$. For example, let us consider a case with $N=4$. The state of the system
is then given by $|+\rangle |+\rangle |+\rangle |+\rangle $. $A$ stores half
of the qubits $Q_{A}$ of the Bell pairs in a quantum smart card. $B$ keeps
the other half $Q_{B}$ of the Bell pairs in a quantum storage device. The
process is depicted in Fig. 1. Box $A$ represents $A$'s smart card and box $%
B $ represents the quantum storage device. The circles connected by wavy
lines represent entangled qubits. (2) $A$ generates an $N$-bit classical
password $K$ composed of $0$s and $1$s, keeping it secret from $B$ and
others. For instance, let us consider $K=(0101)$. $A$ performs a unitary
transformation $R$ on the qubits of $Q_{A}$ corresponding to the bit values $%
1$ of $K$. The action of $R$ changes $|+\rangle $ into another Bell state $%
|\xi \rangle $. In the example, the state of $Q_{A}$ and $Q_{B}$ is
transformed into $|+\rangle |\xi \rangle |+\rangle |\xi \rangle $. The
process of (2) is depicted in Fig. 2. (3) In authentication, $A$ generates a
quantum password $Q_{K}$. The bit values $0$ of $K$ is encoded into a qubit
state $|0\rangle $ and $1$ of $K$ is encoded into a state $|\alpha \rangle $
non-orthogonal to $|0\rangle $. In the example, the quantum password is
generated as $Q_{K}=|0\rangle |\alpha \rangle |0\rangle |\alpha \rangle $. \
(4) $A$ sends $Q_{A}$ and $Q_{K}$ to $B$ through a quantum channel. (5) $B$
unlocks $Q_{A}$ and $Q_{B}$ using a quantum device $UL$ with the quantum
password $Q_{K}$. The device $UL$ operates so as to perform a unitary
transformation $U$ on the composite system of $Q_{K}$, $Q_{A}$ and $Q_{B}$. $%
U$ \ does not change the input state $|0\rangle |+\rangle $ with bit values
\ $0$ of $K$ and transforms the input state $|\alpha \rangle |\xi \rangle $
with bit values $1 $ of $K$ into $|c\rangle |+\rangle $, where $|c\rangle $
is a quantum state of a qubit of $Q_{K}$. Fig. 3 depicts the input state for
the $N=4$ example and Fig. 4 shows the output state. (6) $B$ checks whether
the output state of $Q_{A}$ and $Q_{B}$ is $|+\rangle ^{\otimes N}$. This is
done by a Bell measurement of $|+\rangle $. If a positive result is
obtained, $B$ recognizes $A$. If not, $B$ stops the process. (7) $B$ locks $%
Q_{A}$ and $Q_{B}$ by a quantum device $L$ with $Q_{K}$. The action of $L$
is the inverse transformation of $U$. In the example, the output state of $L$
for $Q_{A}$ and $Q_{B}$ is given by $|+\rangle |\xi \rangle |+\rangle |\xi
\rangle $. The output state of $L$ for $Q_{K}$ is given by $|0\rangle
|\alpha \rangle |0\rangle |\alpha \rangle $. (8) $B$ breaks off $Q_{K}$ and
erases the information so it cannot be stolen by others. (9) $B$ returns $%
Q_{A}$ to $A$ and $A$ restores $Q_{A}$ to the smart card.

\bigskip

We note that $Q_{A}$ and $Q_{B}$ cannot be decoded correctly by $B$ without
the information of $K$ or $Q_{K}$. Therefore, even if the smart card is
stolen by $E$, $E$ cannot impersonate $A$ without $K$. Thus, property (I) in
section 1 is achieved. Moreover, the contracted states of the qubits of $%
Q_{A}$ become the maximal entropy state:

\begin{equation*}
\rho _{\max }=\frac{I}{2}.
\end{equation*}%
Thus property (II) in section 1 is verified. Similarly, the contracted
states of $Q_{B}$ also become the maximal entropy state. Therefore, the
storage device of $B$ does not contain any information of $K$. It is also
stressed that the only information that $B$ holds is $Q_{B}$. Hence $E$
cannot steal useful information about $K$ from $B$'s storage device. This
guarantees property (III). In step (4), the information of $K$ is encoded by
two non-orthogonal states. Thus, $E$ cannot perfectly obtain $Q_{K}$ by
eavesdropping. By using an extension which will be proposed in section 5,
rapid detection of eavesdropping also becomes possible and achieves property
(IV). It is notable that there exists no entanglement between $A$'s quantum
password $Q_{K}$ and the composite system of $Q_{A}$ and $Q_{B}$. Therefore, 
$Q_{K}$ is disposable in each round of the protocol. Therefore property (V)
is attained. $B$ cannot get perfect information of $K$ in steps (5) and (6)
because the quantum qubits accessible by $B$ are all non-orthogonal to each
other.

In the following, we give explicit forms of the unitary transformation $R$
and $U$. Let us define four Bell states orthogonal to each other as follows: 
\begin{align*}
|\pm \rangle & =\frac{1}{\sqrt{2}}\left[ |0\rangle |0\rangle \pm |1\rangle
|1\rangle \right] , \\
|B_{\pm }\rangle & =\frac{1}{\sqrt{2}}\left[ |0\rangle |1\rangle \pm
|1\rangle |0\rangle \right] ,
\end{align*}%
where $|0\rangle $ and $|1\rangle $ are two orthonormal states of a qubit.
The first qubit is stored by $A$ and the second by $B$. The state $|\alpha
\rangle $, which is used when $A$'s quantum password is generated, is given
by

\begin{equation*}
|\alpha \rangle =\alpha |0\rangle +\beta |1\rangle ,
\end{equation*}%
where $\alpha $ and $\beta $ are real constants such that $0\leq \alpha <1$
and $\beta =\sqrt{1-\alpha ^{2}}$. The unitary transformation $R$ is defined
as follows: 
\begin{align*}
R|0\rangle & =e^{i\delta }|0\rangle , \\
R|1\rangle & =e^{-i\delta }|1\rangle ,
\end{align*}%
where $\delta $ is a real parameter. For later convenience, we introduce two
real parameters $\xi $ and $\eta $ such that $e^{i\delta }=\xi +i\eta $, $%
0\leq \xi <1$ and $\eta =\sqrt{1-\xi ^{2}}$. Acting on $A$'s qubit in $%
|+\rangle $ with $R$ yields a new Bell state $|\xi \rangle $:

\begin{equation}
R\otimes I|+\rangle =|\xi \rangle ,  \label{1}
\end{equation}%
where $|\xi \rangle $ is given by

\begin{equation*}
|\xi \rangle =\xi |+\rangle +i\eta |-\rangle .
\end{equation*}%
It is easy to check explicitly that $|\xi \rangle $ is a Bell state because
the following relations hold: 
\begin{align*}
\rho _{A}& =\limfunc{Tr}_{B}\left[ |\xi \rangle \langle \xi |\right] =\frac{I%
}{2}, \\
\rho _{B}& =\limfunc{Tr}_{A}\left[ |\xi \rangle \langle \xi |\right] =\frac{I%
}{2}.
\end{align*}%
Because $|+\rangle $ is also a Bell state, the following relations are also
satisfied:

\begin{align*}
\rho _{A}& =\limfunc{Tr}_{B}\left[ |+\rangle \langle +|\right] =\frac{I}{2},
\\
\rho _{B}& =\limfunc{Tr}_{A}\left[ |+\rangle \langle +\right] =\frac{I}{2}.
\end{align*}%
Consequently no information of $K$ can be extracted from only $Q_{A}~$or $%
Q_{B}$.

Here it should be noted that a similar idea of imprinting information into
Bell states by a local operation has been proposed in \cite{Shi}. However,
the authors treat only orthogonal Bell states. In contrast, non-orthogonal
Bell states play a crucial role in our protocol. Because $|+\rangle $ and $%
|\xi \rangle $ are not orthogonal to each other, $B$ cannot decode $K$
perfectly by a measurement of $Q_{A}$ and $Q_{B}$.

The unitary transformation $U$ is defined such that the following relations
are satisfied:

\begin{align}
U|0\rangle |+\rangle & =|0\rangle |+\rangle ,  \label{2} \\
U|1\rangle |+\rangle & =\frac{\beta }{d}\xi |1\rangle |+\rangle
+u_{11}|0\rangle |-\rangle +u_{21}|1\rangle |-\rangle ,  \notag
\end{align}

\begin{align*}
U|0\rangle |-\rangle & =-i\frac{\alpha \eta }{d}|1\rangle |+\rangle
+u_{12}|0\rangle |-\rangle +u_{22}|1\rangle |-\rangle , \\
U|1\rangle |-\rangle & =-i\frac{\beta \eta }{d}|1\rangle |+\rangle
+u_{13}|0\rangle |-\rangle +u_{23}|1\rangle |-\rangle , \\
U|b\rangle |B_{\pm }\rangle & =|b\rangle |B_{\pm }\rangle ,
\end{align*}%
where $d$ is a real parameter given by $d=\sqrt{1-\alpha ^{2}\xi ^{2}}$ and $%
u_{ij}$ are complex numbers satisfying uniary relations given by%
\begin{align*}
\left\vert u_{11}\right\vert ^{2}+\left\vert u_{21}\right\vert ^{2}& =1-%
\frac{\beta ^{2}\xi ^{2}}{d^{2}}, \\
\left\vert u_{12}\right\vert ^{2}+\left\vert u_{22}\right\vert ^{2}& =1-%
\frac{\alpha ^{2}\eta ^{2}}{d^{2}}, \\
\left\vert u_{13}\right\vert ^{2}+\left\vert u_{23}\right\vert ^{2}& =1-%
\frac{\beta ^{2}\eta ^{2}}{d^{2}},
\end{align*}%
\begin{align*}
u_{11}u_{13}^{\ast }+u_{21}u_{23}^{\ast }& =-i\frac{\beta ^{2}\xi \eta }{%
d^{2}}, \\
u_{12}u_{13}^{\ast }+u_{22}u_{23}^{\ast }& =-\frac{\alpha \beta \eta ^{2}}{%
d^{2}}, \\
u_{11}u_{12}^{\ast }+u_{21}u_{22}^{\ast }& =-i\frac{\alpha \beta \xi \eta }{%
d^{2}}.
\end{align*}
$U$ has the following properties. For bit values $0$ of $K$, the input state 
$|0\rangle |+\rangle $ does not change at all under $U$ operation as seen in
Eq. (\ref{2}). For bit values $1$ of $K$, the input state $|\alpha \rangle
|\xi \rangle $ is transformed into

\begin{equation}
U|\alpha \rangle |\xi \rangle =|c\rangle |+\rangle ,  \label{3}
\end{equation}%
where $|c\rangle $ is a state defined by%
\begin{equation}
|c\rangle =\alpha \xi |0\rangle +d|1\rangle .~  \label{100}
\end{equation}%
Eq. (\ref{3}) can be directly verified from Eq.(\ref{100}) and the inverse
relation $U^{-1}|c\rangle |+\rangle =|\alpha \rangle |\xi \rangle $, which
is derived from unitary relations such that

\begin{eqnarray*}
U^{-1}|0\rangle |+\rangle &=&|0\rangle |+\rangle , \\
U^{-1}|1\rangle |+\rangle &=&\frac{\beta \xi }{d}|1\rangle |+\rangle +i\frac{%
\alpha \eta }{d}|0\rangle |-\rangle +i\frac{\beta \eta }{d}|1\rangle
|-\rangle .
\end{eqnarray*}%
From Eq. (\ref{2}) and Eq. (\ref{3}), it is verified that entanglement
between $Q_{K}~$and $Q_{A}+Q_{B}$ is not generated before and after the
operation of $U$ and $U^{-1}$. Therefore, purity of the state for $%
Q_{A}+Q_{B}$ is preserved even if $Q_{K}$ is discarded by $B$ after the
authentication. This fact allows us to repeat the use of $Q_{A}$ stored in
the smart card.

\bigskip

\section{Security Analysis}

\bigskip

In this section, we present security analysis of the above protocol. First,
we assume that $E$ does not have $A$'s smart card and her password $K$. The
success probability $p_{s}$ of $E$ per qubit to pass the authentication test
by $B$ is evaluated as follows. Without access to $K$, $E$ has to prepare a
universal optimal state $|\Psi _{E}\rangle ^{\otimes N}$ of $Q_{K}+Q_{A}$ as
a forged quantum password and a forged smart card. Without loss of
generality, $|\Psi _{E}\rangle $ is written as

\begin{equation*}
|\Psi _{E}\rangle =\Psi _{00}|0\rangle |0\rangle +\Psi _{01}|0\rangle
|1\rangle +\Psi _{10}|1\rangle |0\rangle +\Psi _{11}|1\rangle |1\rangle ,
\end{equation*}%
where $\Psi _{bb^{\prime }}$ are complex coefficients satisfying the
normalization condition of the state. The first qubit corresponds to $Q_{K}$
and the second to $Q_{A}$. Because the forged input of $E$ is not at all
entangled with $Q_{B}$, the input state of $Q_{B}$ of $U$ is given by $%
I_{B}/2$, independent of the bit values of $K$. Hence $p_{s}$ is written as

\begin{eqnarray*}
p_{s} &=&\langle +|\limfunc{Tr}_{K}\left[ U\left( |\Psi _{E}\rangle \langle
\Psi _{E}|\otimes \frac{I_{B}}{2}\right) U^{\dag }\right] |+\rangle \\
&=&\frac{1}{2}\left\vert \langle 0|\langle +|U|\Psi _{E}\rangle |0\rangle
\right\vert ^{2}+\frac{1}{2}\left\vert \langle 0|\langle +|U|\Psi
_{E}\rangle |1\rangle \right\vert ^{2} \\
&&+\frac{1}{2}\left\vert \langle 1|\langle +|U|\Psi _{E}\rangle |0\rangle
\right\vert ^{2}+\frac{1}{2}\left\vert \langle 1|\langle +|U|\Psi
_{E}\rangle |1\rangle \right\vert ^{2},
\end{eqnarray*}%
where the trace is taken in terms of the state space of $Q_{K}$. Through
straightforward manipulations, $p_{s}$ is evaluated as

\begin{align*}
p_{s}& =\frac{1}{4}\left\vert \Psi _{00}\right\vert ^{2}+\frac{1}{4}%
\left\vert \Psi _{01}\right\vert ^{2} \\
& +\frac{1}{4}\left\vert \frac{\beta }{d}e^{-i\delta }\Psi _{10}-i\frac{%
\alpha \eta }{d}\Psi _{00}\right\vert ^{2} \\
& +\frac{1}{4}\left\vert \frac{\beta }{d}e^{i\delta }\Psi _{11}+i\frac{%
\alpha \eta }{d}\Psi _{01}\right\vert ^{2},
\end{align*}%
by use of the following unitary relations.

\begin{eqnarray*}
\langle 0|\langle +|U &=&\langle 0|\langle +|, \\
\langle 1|\langle +|U &=&\frac{\beta \xi }{d}\langle 1|\langle +|-i\frac{%
\alpha \eta }{d}\langle 0|\langle -|-i\frac{\beta \eta }{d}\langle 1|\langle
-|.
\end{eqnarray*}%
We can show that $p_{s}$ is bounded above by applying Schwartz inequalities
as follows: 
\begin{align*}
p_{s}& \leq \frac{1}{4}\left\vert \Psi _{00}\right\vert ^{2}+\frac{1}{4}%
\left\vert \Psi _{10}\right\vert ^{2} \\
& +\frac{1}{4}\left( \left\vert \Psi _{10}\right\vert ^{2}+\left\vert \Psi
_{00}\right\vert ^{2}\right) \\
& +\frac{1}{4}\left( \left\vert \Psi _{11}\right\vert ^{2}+\left\vert \Psi
_{01}\right\vert ^{2}\right) \\
& =\frac{1}{4}\left( \left\vert \Psi _{00}\right\vert ^{2}+\left\vert \Psi
_{10}\right\vert ^{2}\right) +\frac{1}{4} \\
& \leq \frac{1}{2}.
\end{align*}%
Consequently, we get a lower bound of detection probability of $E$'s
impersonation as

\begin{equation*}
p_{E}^{(N)}\geq 1-\left( \frac{1}{2}\right) ^{N}.
\end{equation*}%
Therefore, $B$ is able to detect $E$ with a high probability for a large $N$.

Next, we consider a case where $E$ succeeds in stealing $A$'s card. However,
we assume that $E$ does not know $K$. In this case, $E$ must make a forged
quantum password. The optimal state for each qubit is denoted by $\rho
_{E}=|K_{E}\rangle \langle K_{E}|$. \ Assuming $K$ is randomly generated, \
the appearance probability of each bit value of $K$ is $1/2$. For bit values 
$0$ of $K$, the state of $Q_{A}+Q_{B}$ is $|+\rangle $. Thus, the detection
probability of $E$ per qubit is given by 
\begin{equation*}
p_{E0}=\limfunc{Tr}\left[ U\left( \rho _{E}\otimes |+\rangle \langle
+|\right) U^{\dag }\left( I-|+\rangle \langle +|\right) \right] .
\end{equation*}%
\ For bit values $1$, the state is $|\xi \rangle $. Hence, the detection
probability is written as

\begin{equation*}
p_{E1}=\limfunc{Tr}\left[ U\left( \rho _{E}\otimes |\xi \rangle \langle \xi
|\right) U^{\dag }\left( I-|+\rangle \langle +|\right) \right] .
\end{equation*}%
The average detection probability per qubit is given by%
\begin{equation*}
\Delta _{E}=\frac{1}{2}p_{E0}+\frac{1}{2}p_{E1}.
\end{equation*}%
We parametrize $\rho _{E}$ as

\begin{equation*}
\rho _{E}=r|0\rangle \langle 0|+(1-r)|1\rangle \langle 1|+\left( x+iy\right)
|0\rangle \langle 1|+\left( x-iy\right) |1\rangle \langle 0|,
\end{equation*}%
where

\begin{equation*}
0\leq r\leq 1,x^{2}+y^{2}=r(1-r).
\end{equation*}%
The detection probability is then evaluated as

\begin{equation*}
\Delta _{E}=\frac{1}{2}\frac{1-\xi ^{2}}{1-\alpha ^{2}\xi ^{2}}\left[
1+\alpha ^{2}-2\alpha ^{2}r-2\alpha \beta x\right] .
\end{equation*}%
The minimum value of $\Delta _{E}$ in terms of $r$ and $x$ is easily
obtained as

\begin{equation*}
p_{n}=\min \Delta _{E}=\frac{1}{2}\frac{\left( 1-\xi ^{2}\right) }{1-\alpha
^{2}\xi ^{2}}\left( 1-\alpha \right) .
\end{equation*}%
For instance, taking typical values of $\alpha $ and $\xi $ as $\alpha =\xi
=1/2$, the value of $p_{n}$ is evaluated as $p_{n}=1/5$. It is noted that
the total detection probability of $E$ is given by%
\begin{equation*}
p_{n}^{(N)}=1-(1-p_{n})^{N}.
\end{equation*}%
Consequently, $E$ can be detected with a high rate for a large $N$.

A comment should be made about man-in-the-middle attacks. If $E$ is able to
secretly occupy classical and quantum channels between $A$ and $B$ and
perform any attack allowed by physics laws, $E$ is able to steal all the
quantum states of $A$ in the transfer through the channels. In order to
impersonate $A$ after this round of the protocol, $E$ must keep the stolen
qubits. However, we have assumed that $E$ cannot prevent $B$ from knowing
the start of $A$'s authentication protocol through a public channel. In
order to avoid $B$ quickly noticing the impersonation, $E$ has to send some
forged qubits as $Q_{A}$ and $Q_{K}$ to $B$. Then the identification test by 
$B$ yields a wrong output and the man-in-the-middle attack is easily noticed.

Though our protocol has many advantages, as detailed above, some subtle
loopholes exist. One of them may occur in step (4). Because of the
non-orthogonality of \ $|0\rangle $ and $|\alpha \rangle $, perfect cloning
of $Q_{K}$ in the channel is prohibited. However, it is possible for $E$ to
attempt an approximate cloning of $Q_{K}$. Even though the cloning leaves a
disturbance in the states of $Q_{K}$ received by $B$, the detection rate of
eavesdropping is not large. If $B$ fails to detect $E$, $E$ may next try to
steal $A$'s smart card. Let us assume that $E$ succeeds in obtaining the
card. This card-steal attack with an approximate clone of $Q_{K}$ strongly
decreases $B$'s probability of detecting impersonation. Another loophole may
occur in step (8). $B$ discards $A$'s quantum password $Q_{K}$ after the
authentication. If $E$ infiltrates $B$'s region and secretly measures $Q_{K}$
and accumulates the results, $E$ can estimate $K$ with high precision after
several rounds by $A$ and $B$. This may lead to possible abuse of the
information by $E$. In the next section, we give an improved protocol robust
against these attacks, without loss of the advantages of the basic protocol.

\section{Extended protocol}

\bigskip

In this section, we present an extended protocol, which increases the
detection probability of eavesdropping in quantum channels and decreases the
amount of knowledge of $K$ leaked in the discarding step of $Q_{K}$. Before $%
A$ sends $Q_{K}$ to $B$, $A$ generates a long sequence of two non-orthogonal
states of qubits like the BB84 quantum key distribution \cite{BB84} and uses
them as decoys to detect eavesdropping. In order to suppress the leakage of
information about $K$, a one-time-pad method is adopted when $A$ generates a
quantum password. The extended protocol is composed of 14 steps as follows.

\bigskip

(1) $A$ and $B$ meet and generate $N$ \ qubit pairs in a Bell state $%
|+\rangle $. $A$ stores half of the Bell pairs $Q_{A}$ in a quantum smart
card. $B$ keeps the other half $Q_{B}$ in a quantum storage device. (2) $A$
generates an $N$-bit classical password $K$ composed of $0$s and $1$s and
keeps it secret from $B$ and others. $A$ performs $R$ for $Q_{A}$ qubits
corresponding to bit values $1$ of $K$. (3) In authentication, the lock
process is reversed using $K$. $A$ performs $R^{-1}$ for $Q_{A}$ qubits
corresponding to bit values $1$ of $K$. (4) $A$ generates $N$-bit
pseudo-random numbers $\tilde{K}$. \thinspace $\tilde{K}$ is used as a
one-time-pad password in the transfer of $A$'s qubits. $A$ performs $R$ for $%
Q_{A}$ corresponding to bit values $1$ of $\tilde{K}$. (5) $A$ generates a
quantum password $Q_{\tilde{K}}$. The bit values $0$ of $\tilde{K}$ are
encoded into a qubit state $|0\rangle $ and bit values $1$ of $\tilde{K}$
into $|\alpha \rangle $. \ (6) $A$ generates a sequence of $N_{D}$
pseudo-random numbers $K_{D}$ composed of $2,3,4,5$. $K_{D}$ is quantum
mechanically encoded using four quantum states $|0\rangle ,|1\rangle $ and
\bigskip 
\begin{align*}
|0_{\times }\rangle & =\frac{1}{\sqrt{2}}\left( |0\rangle +|1\rangle \right)
, \\
|1_{\times }\rangle & =\frac{1}{\sqrt{2}}\left( |0\rangle -|1\rangle \right)
.
\end{align*}%
The number \thinspace $2$ in $K_{D}$ is replaced by $|0\rangle $, 3 by $%
|1\rangle $, 4 by $|0_{\times }\rangle $ and 5 by $|1_{\times }\rangle $. We
call the sequence of qubits $Q_{D}$. (7) $A~$makes $Q_{\tilde{K}}$ randomly
slip into $Q_{D}$ and sends $Q_{A}$ and $Q_{\tilde{K}}+Q_{D}$ to $B$. (8)
After they are received by $B$, $A$ announces to $B$ the positions of the
qubits of $Q_{D}$ \ and the value of $K_{D}$. $B$ then separates $Q_{D}$
from $Q_{\tilde{K}}$. $B$ measures $Q_{D}$ in the basis of $\left\{
|0\rangle ,|1\rangle \right\} $ if the value of $K_{D}$ is $2$ or $3\,$\ and
in the basis of $\left\{ |0_{\times }\rangle ,|1_{\times }\rangle \right\} $
if the value of $K_{D}$ is $4$ or $5$. If the results are consistent with $%
K_{D}$, $B$ makes a judgement that there is no eavesdropping. If it is
judged that eavesdropping may have occurred, $B$ stops the process. (9) $B$
unlocks $Q_{A}$ and $Q_{B}$ by the unitary transformation $U$ with the
password $Q_{\tilde{K}}$. $U$ \ does not change the input state $|0\rangle
|+\rangle $ with bit value \ $0$ of $\tilde{K}$ and transforms the input
state $|\alpha \rangle |\xi \rangle $ with bit value $1$ of $\tilde{K}$ into 
$|c\rangle |+\rangle $. (10) $B$ checks whether the output state of $%
Q_{A}+Q_{B}$ is $|+\rangle ^{\otimes N}$. If the result is positive, $B$
recognizes $A$. If not, $B$ stops the process. (11) $B$ locks $Q_{A}$ and $%
Q_{B}$ by $U^{-1\text{ }}$with $Q_{\tilde{K}}$. (12) $B$ discards $Q_{\tilde{%
K}}$. (13) $B$ returns $Q_{A}$ to $A$ and $A$ restores $Q_{A}$ to the smart
card. (14) $A$ performs $R^{-1}$ to the qubits of $Q_{A}$ corresponding to
bit values $1$ of $\tilde{K}$ to make the state of $Q_{A}+Q_{B}$ to be $%
|+\rangle ^{\otimes N}$. Then, $A$ locks $Q_{A}+Q_{B}$ by the original
password $K$. $A$ performs $R$ on the qubits of $Q_{A}$ corresponding to bit
values $1$ of $K$.

The one-time-pad password method in step (4) prevents $E$ from stealing the
information of the original password $K$ when the quantum password is
discarded by $B$. Step (8) of the protocol also prevents approximate cloning
attacks by $E$ eavesdropping by taking a large $N_{D}$. A detailed security
analysis will be reported elsewhere.

It is expected that the protocol will protect the basic infrastructure of
the information-based society if a quantum smart card is devised which can
store quantum information for a long period.

\bigskip

\textbf{Acknowledgement}\newline

This research was partially supported by the SCOPE project of the MIC.

\bigskip

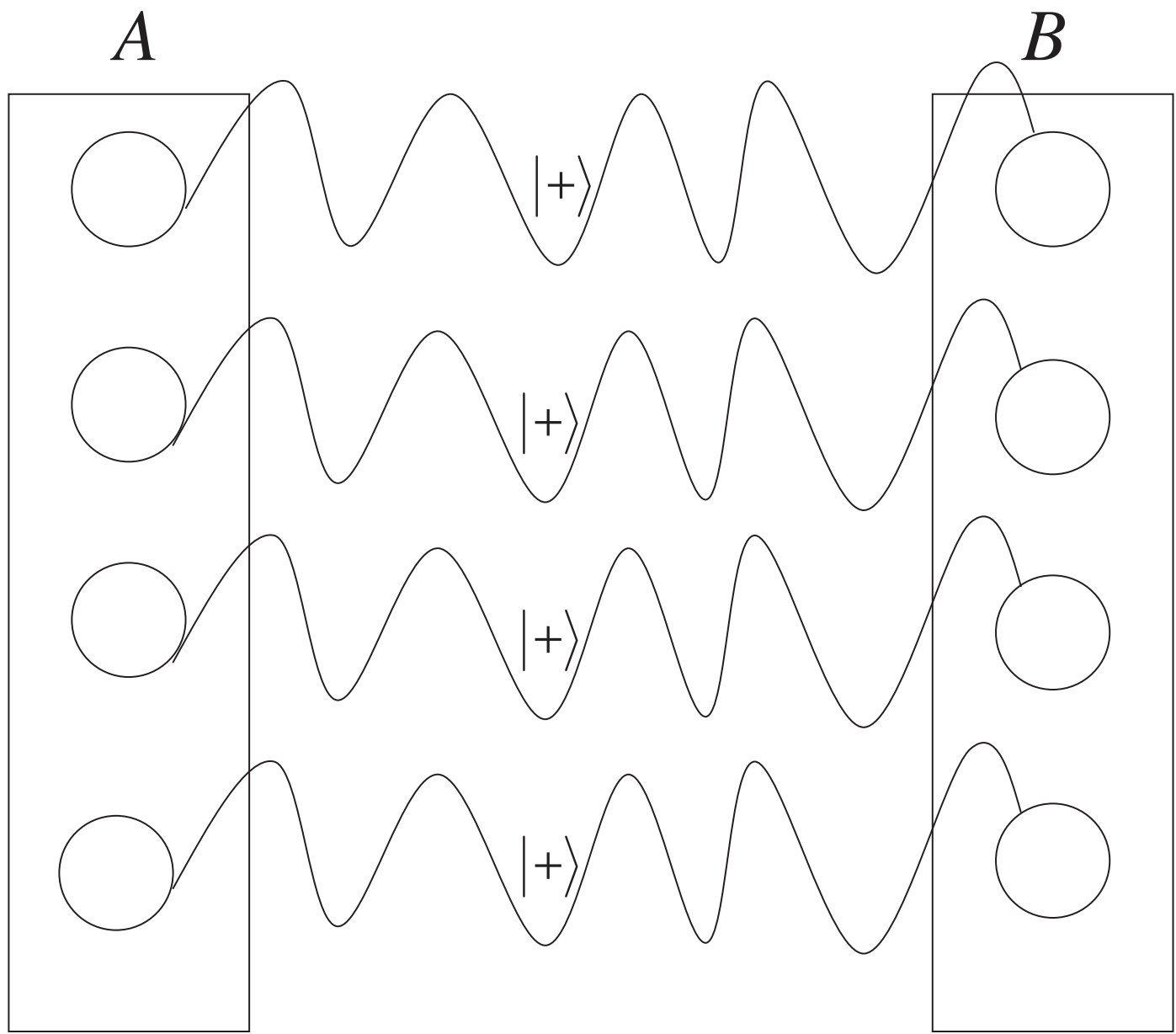
Fig. 1: In the basic protocol, $A$ and $B$ meet and share Bell pairs. The
case with $K=(0101)$ is depicted. The state of the system is given by $%
|+\rangle |+\rangle |+\rangle |+\rangle $. $A$ stores half of the Bell pairs 
$Q_{A}$ in a quantum smart card. $B$ keeps the other half $Q_{B}$ in a
quantum storage device. Box $A$ indicates the smart card and box $B$
indicates the quantum storage device. The circles connected by wavy lines
represent entangled qubits.

\bigskip

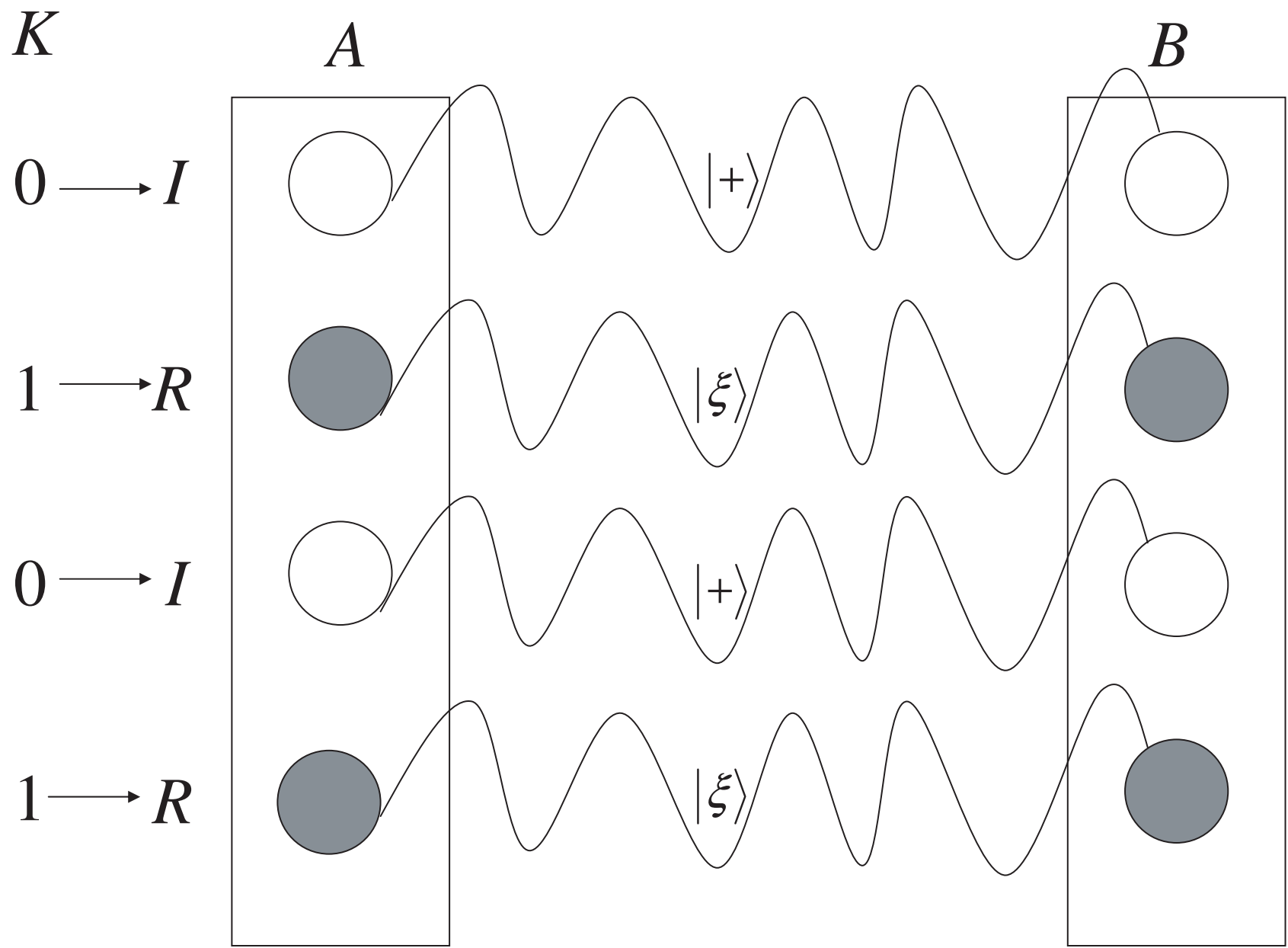
Fig. 2: $A$ generates a classical password $K$ composed of $0$s and $1$s and
keeps it secret, even from $B$. The case with $K=(0101)$ is depicted. $A$
performs a unitary transformation $R$ on the qubits of $Q_{A}$ corresponding
to bit values $1$ of $K$. The action of $R$ changes $|+\rangle $ into
another Bell state $|\xi \rangle $. In this example, the state of $Q_{A}$
and $Q_{B}$ is given by $|+\rangle |\xi \rangle |+\rangle |\xi \rangle $.

\bigskip

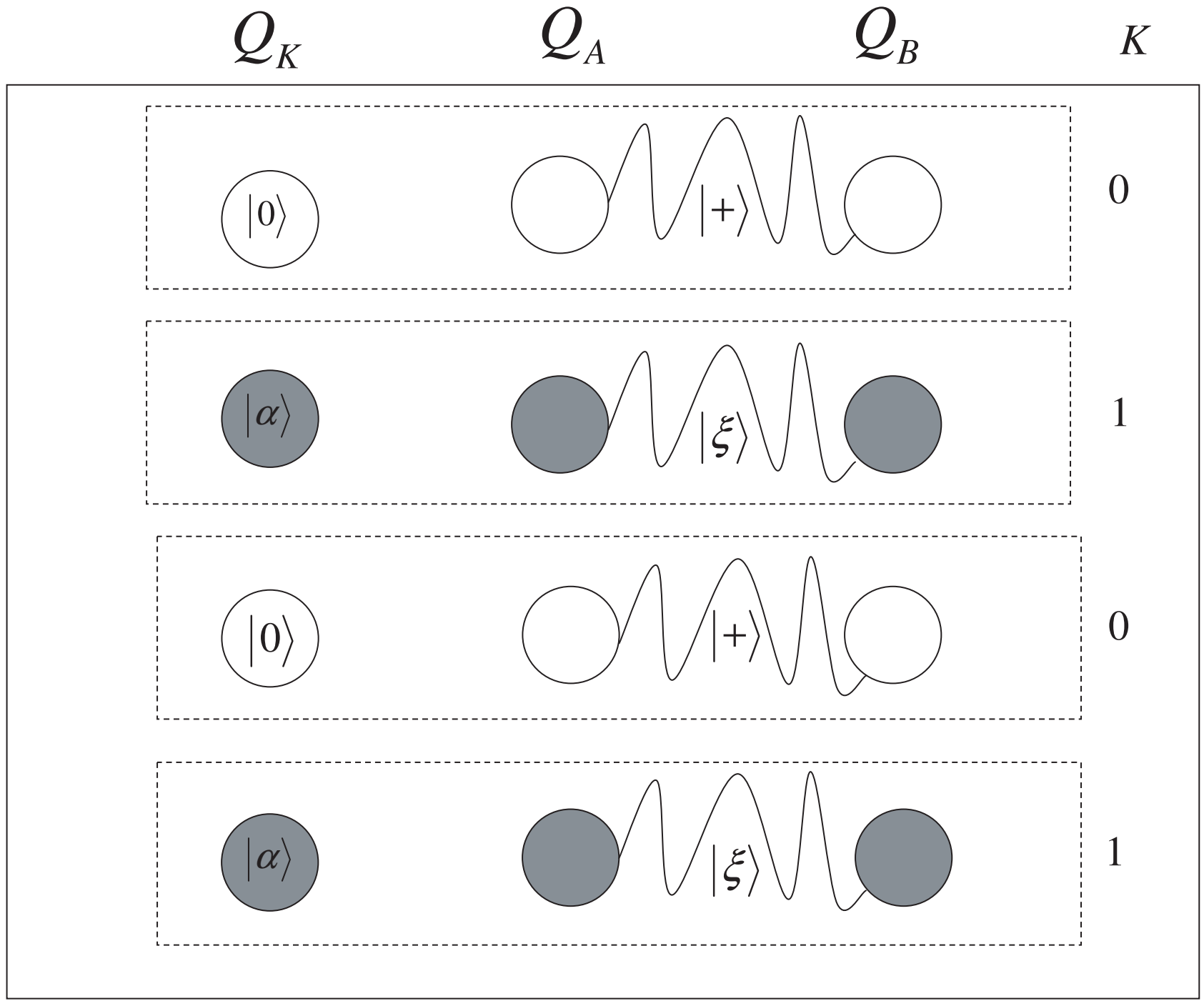
Fig. 3: $B$ unlocks $Q_{A}$ and $Q_{B}$ using a quantum device $UL$ with the
password $Q_{K}$. The device $UL$ operates so as to perform a unitary
transformation $U$ on the composite system of $Q_{K}$, $Q_{A}$ and $Q_{B}$.
The input state is depicted for the example.

\bigskip

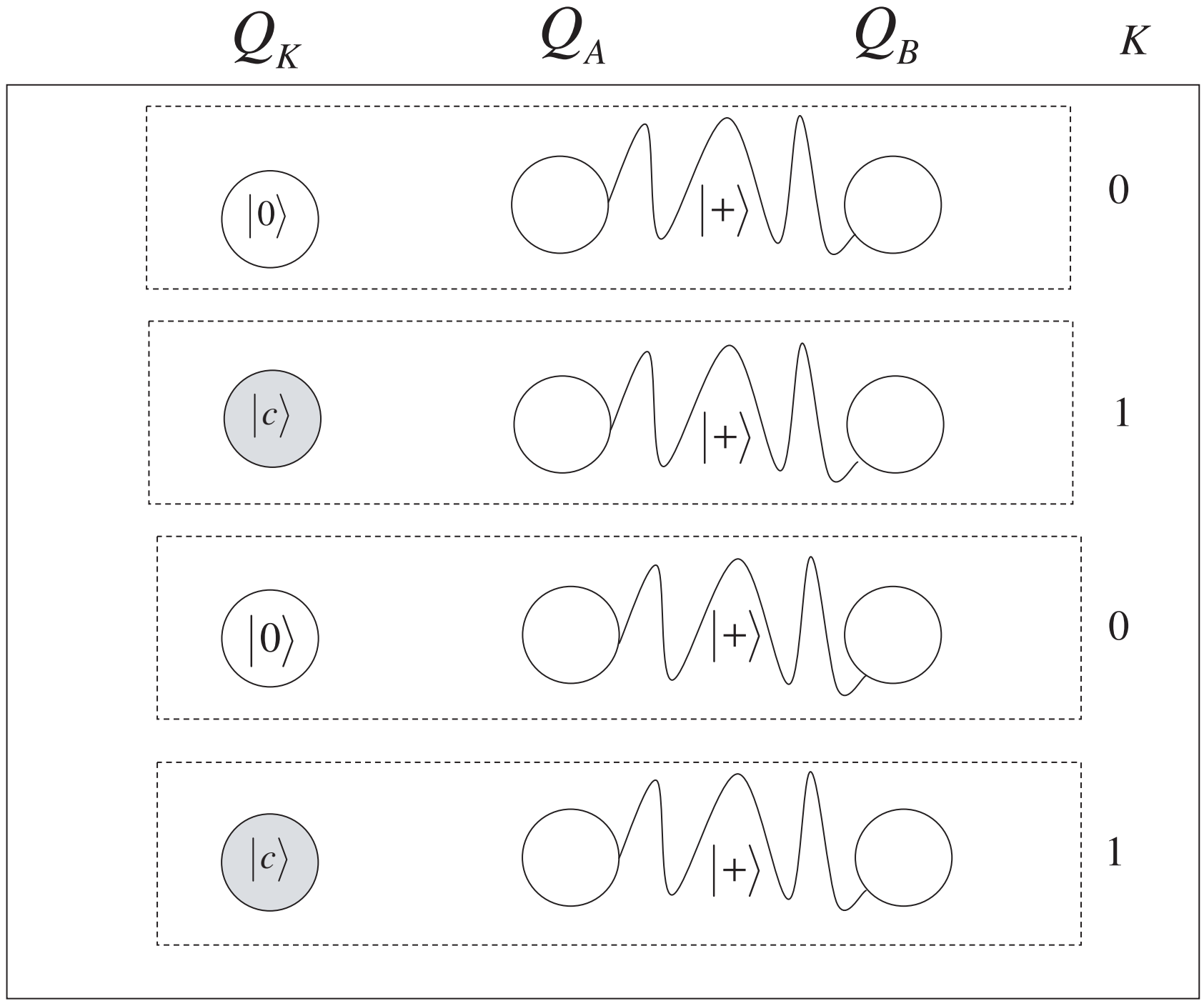
Fig. 4: The output state of $UL$ is depicted.

\end{document}